\documentclass{webofc}
\usepackage[varg]{txfonts}   
\newcommand{\om}{\omega}
\newcommand{\order}{\mathcal{O}}
\newcommand{\braket}[1]{\left\langle#1\right\rangle}
\newcommand{\mrm}{\mathrm}

\begin{document}
\title{QCD at high temperature and density: selected highlights}

\author{\firstname{Jon-Ivar} \lastname{Skullerud}\inst{1}\fnsep\thanks{\email{jonivar.skullerud@mu.ie}}
}

\institute{Department of Theoretical Physics, Maynooth University,
  Maynooth, Co Kildare, Ireland}

\abstract{%
I review some of the recent progress in QCD at high temperature and
density, with a focus on the nature of the high-temperature
transition; cold and dense matter; and hadron properties and transport
coefficients at high temperature.
}
\maketitle
\section{Introduction}
\label{intro}

Understanding the phase diagram of QCD and the properties of hot and
dense strongly interacting matter is an endeavour of great scope
and interest.  Besides the theoretical interest in the exotic forms of
matter that may appear under these conditions, a full understanding
will cast light on the formation of matter in the early universe, the
properties of neutron stars, the formation of heavy elements, and much
more.  Aligned to this is the experimental programme at RHIC and LHC
and, in the near future, at FAIR and NICA.

The theoretical effort that has gone into this field over the past
number of years is so large and varied that it will not be possible to
do it full justice in one talk, and this overview is therefore
necessarily selective.  Important topics that will not be covered here
include the QCD equation of state at high temperature, QCD in nonzero
magnetic field, topology and deconfinement, non-equilibrium dynamics,
and much more (see for example
\cite{Fukushima:2013rx,Ding:2017giu,Fischer:2018sdj,Ghiglieri:2020dpq,Guenther:2020jwe}
for complementary overviews).  I will instead focus on aspects of the
transition(s) at zero chemical potential $\mu$; exploring the chiral
critical line at nonzero $\mu$; some progress in understanding matter
at large $\mu$ and low temperature $T$; and hadron properties and
transport coefficients at large $T$.  The emphasis will be on lattice
QCD results, but not exclusively so: in particular, results using
other Dyson--Schwinger equations (DSEs), the functional
renormalisation group (FRG) and perturbation theory will be
referenced where appropriate.

\section{Transition(s) at zero chemical potential}
\label{sec:transitions}

\subsection{Location and order of the chiral transiton}
\label{sec:Tpc}

\begin{figure}[t]
\centering
\includegraphics[width=0.5\textwidth,clip]{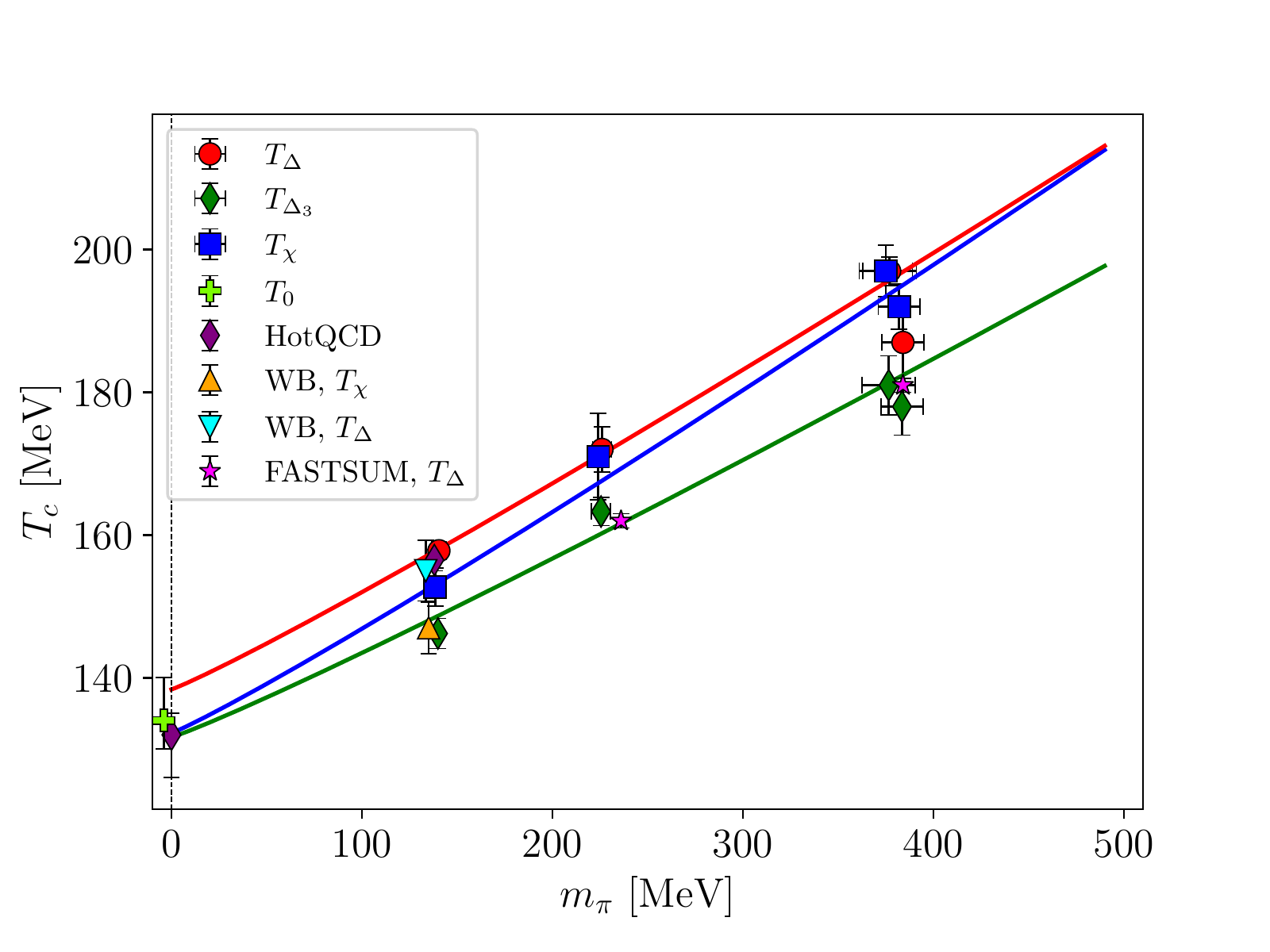}
\caption{Pseudocritical temperature $T_c$ as a function of the pion
  mass $m_\pi$ from different lattice collaborations, from \cite{Kotov:2021rah}.}
\label{fig:Tc}       
\end{figure}

It is by now well established that the transition from the hadronic
phase to the quark--gluon plasma at $\mu=0$ is a crossover,
characterised by a rapid change in the chiral condensate and a peak in
the chiral susceptibility.  Away from the chiral limit (where a true
phase transition is expected) there is therefore no unique definiton
of the pseudocritical temperature $T_c$, but different definitions
(e.g., the peak in the susceptibility $T_\chi$ or the inflection point
of the renormalised chiral condensate $T_\Delta$) give compatible
results for physical quark masses.  There is also good agreement on
the value of $T_c$ at the physical point between lattice groups using
different fermion formulations
\cite{HotQCD:2018pds,Borsanyi:2020fev,Kotov:2021rah} as well as with
recent results from the DSE and FRG approaches
\cite{Gao:2020fbl,Braun:2020ada}, with all finding
$T_\chi=153-158\,$MeV.  Figure~\ref{fig:Tc} \cite{Kotov:2021rah} shows the mass dependence of
$T_c$ from different lattice calculations
\cite{HotQCD:2018pds,Borsanyi:2020fev,Aarts:2020vyb,Kotov:2021rah}
together with scaling functions, illustrating significant progress on
the dependence of $T_c$ on the light quark mass
\cite{HotQCD:2019xnw,Braun:2020ada,Kotov:2021rah}, leading to a
critical temperature in the chiral limit $T_0=130-140\,$MeV.

The order of the chiral phase transition in the massless 2-flavour
limit $m_u,m_d\to0$ is intimately connected with the fate of the
anomalous $U(1)_A$ symmetry.  This symmetry is always broken by
quantum effects, but if the breaking is small (\emph{effective
  restoration}) at $T_c$ the chiral transition may be first order or
second order in the $U(2)_L\otimes U(2)_R/U(2)_V$ universality class.
On the other hand, if the symmetry remains strongly broken, the chiral
transition will be second order in the $O(4)$ universality class.
Lattice studies have yet to arrive at a consensus on this issue, with
\cite{HotQCD:2012vvd,Buchoff:2013nra,Dick:2015twa,Ding:2020xlj,Kaczmarek:2021ser}
finding that the $U(1)_A$ symmetry remains broken, while
\cite{Cossu:2013uua,Brandt:2016daq,Tomiya:2016jwr,Aoki:2020noz} finds
that it is restored.  It is worth noting that of these studies, only
\cite{Ding:2020xlj} has performed a continuum extrapolation.
A recent FRG calculation \cite{Braun:2020mhk} is consistent with $U(1)_A$
restoration and an $O(4)$ universality class.  

The chiral transition is expected to be first order in the 3-flavour
massless limit, $m_u=m_d=m_s=0$.  An open question is whether the
transition is first or second order for physical strange quarks, and
if it is first order, what is the critical pion mass $m_\pi^c$
separating the first order transition region from the crossover
region?  The evidence from a number of lattice studies is that the
first order region shrinks as the continuum limit is approached, with
upper limits of $m_\pi^c\lesssim50\,$MeV from highly improved staggered fermions
\cite{Bazavov:2017xul} and $m_\pi^c\lesssim110\,$MeV from
Wilson--clover fermions \cite{Kuramashi:2020meg}.  A recent study with
Wilson--clover fermions \cite{Ohno:lat21} found evidence of a first
order transition, while no such evidence was found in a study using
staggered fermions \cite{Sharma:lat21}.

This question has also been addressed by studying systems with
$N_f>3$, where the first order transition in the chiral limit is
expected to be stronger and persist to larger quark masses.  
Ohno \textit{et al} \cite{Ohno:lat21} found evidence for a first order
transition also for $N_f=4$ using Wilson--clover fermions.  However,
Cuteri \textit{et al.} \cite{Cuteri:2021ikv}, performing a tricritical
scaling analysis on staggered fermion data, found that the transition
is consistent with second order.  A reanalysis of Wilson--clover data
gave the same result.

\subsection{Some surprises}
\label{sec:surprises}

Recently, some intriguing evidence has emerged suggesting that the
phase structure of QCD may be more complicated than previously
thought, even at $\mu=0$.  In
\cite{Rohrhofer:2017grg,Rohrhofer:2019qwq,Rohrhofer:2019qal} it was found that in the
region $T_c<T\lesssim3T_c$, meson correlators exhibit certain
degeneracies that are not present in the non-interacting case.  This
has been explained in terms of a chiral--spin symmetry
SU(2)$_{CS}$ and its flavour extension $\mathrm{SU}(2N_f)$, which is a
symmetry of the chromoelectric interaction, but is broken by the
chromomagnetic interaction.  Following this, it has been conjectured
that the intermediate-temperature regime is a ``stringy
fluid'' where chiral symmetry is restored and the chromomagnetic
interaction is suppressed, but the confining chromoelectric field is
still present.

Independently, it was found \cite{Cardinali:2021mfh} that a
condensate of thermal monopoles forms at roughly the same temperature
in full ($N_f\!=\!2\!+\!1$) QCD as in pure gauge theory, where this coincides
with the deconfinement transition, i.e., at roughly twice the chiral
transition temperature.
Furthermore, Alexandru and Horvath
\cite{Alexandru:2019gdm,Alexandru:2021pap} found an infrared scaling
of the Dirac spectral density, $\rho(\lambda)\propto1/\lambda$, which
only set in at $T>T_{\text{IR}}=200-250\,\mrm{MeV}>T_c$.  They
conjecture that this scaling is characteristic of a strongly coupled
r\'egime which only ends at a much higher (yet to be determined)
temperature $T_{\text{UV}}$.

It is worth noting that the temperature scales for the suggested additional
transitions are different, and that no
connection between the different phenomena has been established.
However, these results do point to the need for further investigations
into the properties of QCD matter above $T_c$.


\section{Chiral transition line}
\label{sec:chiral-T-mu}

\begin{figure}
\centering
\includegraphics[width=0.5\textwidth]{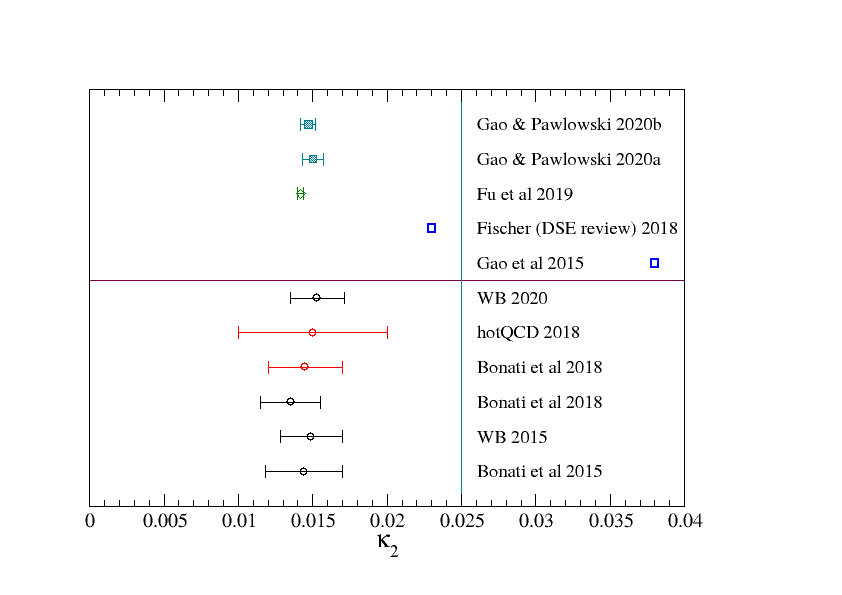}
\includegraphics[width=0.48\textwidth]{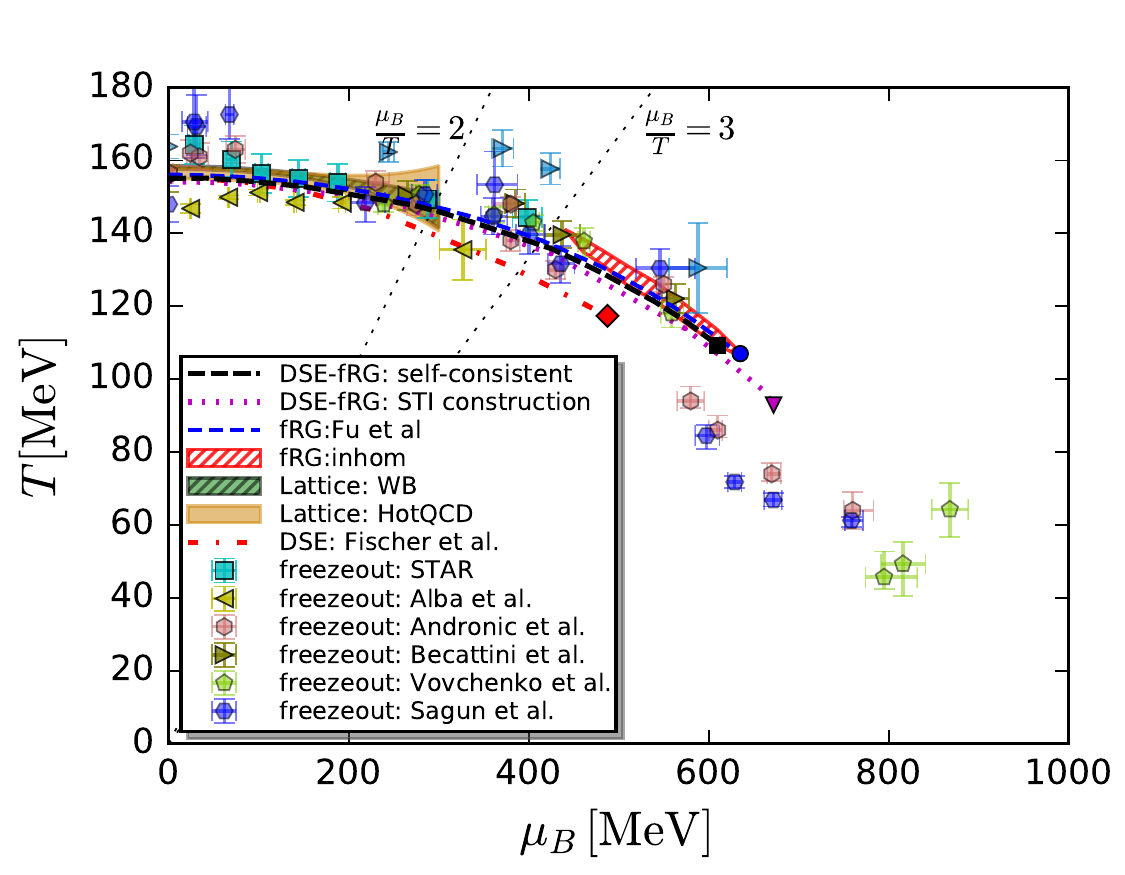}
\caption{Left: The curvature parameter $\kappa_2$ from lattice QCD
  \cite{Bonati:2015bha,Bellwied:2015rza,Bonati:2018nut,HotQCD:2018pds,Borsanyi:2020fev}
  (bottom panel) and functional methods (blue: DSE, green: FRG)
  \cite{Gao:2016qkh,Fischer:2018sdj,Fu:2019hdw,Gao:2020qsj,Gao:2020fbl}
  (top panel). Black points have been obtained by analytic
  continuation from imaginary $\mu$, while red points are from Taylor
  expansion.  Right: Transition line and critical endpoint from
  lattice and functional methods together with experimental freezeout
  lines, from \cite{Gao:2020fbl}.}
\label{fig:curvature}
\end{figure}

The shape of the transition line away from $\mu=0$ has been computed
on the lattice using continuation from imaginary $\mu$
\cite{Bonati:2018nut,Borsanyi:2020fev} and Taylor expansion
\cite{Bonati:2018nut,HotQCD:2018pds} (see also \cite{Borsanyi:2021hbk}
for recent results using a novel reweighting method)
as well as from Dyson--Schwinger equations \cite{Fischer:2018sdj}, the
functional renormalisation group \cite{Fu:2019hdw} and a combination
of the two \cite{Gao:2020qsj,Gao:2020fbl}.  The pseudocritical
temperature $T_c(\mu_B)$ can be expanded in powers of $\mu_B$,
\begin{equation}
\frac{T_c(\mu_B)}{T_c(0)}
    = 1 - \kappa_2\Big(\frac{\mu_B}{T_c}\Big)^2
    - \kappa_4\Big(\frac{\mu_B}{T_c}\Big)^4
    + \order(\mu_B^6)
    \label{eq:Tc-mu}
\end{equation}
leaving the parameters $\kappa_2, \kappa_4$ and any higher order
parameters to be determined.  All the calculations have found that
$\kappa_4$ is consistent with zero within errors.  Results for
$\kappa_2$ are shown in figure~\ref{fig:curvature}.
All the lattice results shown have been continuum extrapolated and are
consistent with each other.  The most recent results from functional
methods are also seen to be converging on the same values.

The right panel of figure~\ref{fig:curvature} shows a summary
\cite{Gao:2020fbl} of lattice, functional and experimental data for
the chiral transition and freezout curves.  In \cite{Gao:2020fbl}
their DSE/FRG calculations were found to be reliable up to
$\mu_B/T=3$, which means that the location of the critical endpoints
shown in the figure should only be considered indicative, and the
spread between the different results reflect the remaining systematic
uncertainties in the approaches.

Fluctuations and correlations of the conserved quantum numbers (baryon
number $B$, electric charge $Q$ and strangeness $S$) play a crucial
r\^ole in connecting theoretical predictions with experimental
results.  From the generalised susceptibilities,
\begin{equation}
  \chi^{B,Q,S}_{i,j,k}
  = \frac{\partial^{i+j+k}(p/T^4)}
  {(\partial\hat{\mu}_B)^i(\partial\hat{\mu}_Q)^i(\partial\hat{\mu}_S)^k}\,,
  \qquad \hat{\mu}_a \equiv \frac{\mu_a}{T}\,
\end{equation}
we may compute experimentally observable quantities such as the mean
$M_B= \chi_1^B$, variance $\sigma_B^2=\chi_2^B$, skewness
$S_B=\chi_3^B/(\chi_2^B)^{3/2}$ and kurtosis $\kappa_B=
\chi^B_4/(\chi^B_2)^{2}$ of the baryon number distribution.  Cumulants
up to 8th order have been computed by the Wuppertal--Budapest
collaboration \cite{Borsanyi:2018grb} and the HotQCD collaboration
\cite{Bazavov:2020bjn} and used to determine skewness and kurtosis at
nonzero baryon chemical potential as well as preliminary results for
the hyper-kurtosis (the sixth moment of the distribution).  Results
have also been obtained in the FRG framework using a low-energy
effective theory \cite{Fu:2016tey,Fu:2021oaw}.  These results may
in turn be used to extract freezeout parameters from experimental
data and check whether experimental results are consistent with
equilibrium QCD.


\section{Large density}
\label{sec:large-mu}

The methods employed to map out the phase diagram at high $T$ and
small $\mu$ break down at lower temperature, in particular in the
region relevant to nuclear matter and the interior of neutron stars.
Recent progress has been made with several alternative methods
including complex Langevin simulations, Lefschetz thimble-based
methods \cite{Alexandru:2020wrj}, density of states
\cite{Langfeld:2016kty,Gattringer:2019khb} and strong coupling
expansions \cite{Philipsen:2021vgp}.  In addition, QCD-like theories
where the sign problem is absent or weak may serve as a laboratory for
generic features of gauge theories at high density as well as provide
checks on the validity of other methods.  Finally, perturbation
theory, valid at asymptotically high densities, can provide
constraints on physics at intermediate densities.  In the following
only a few selected highlights will be presented.

\begin{figure}
\includegraphics[width=0.5\textwidth]{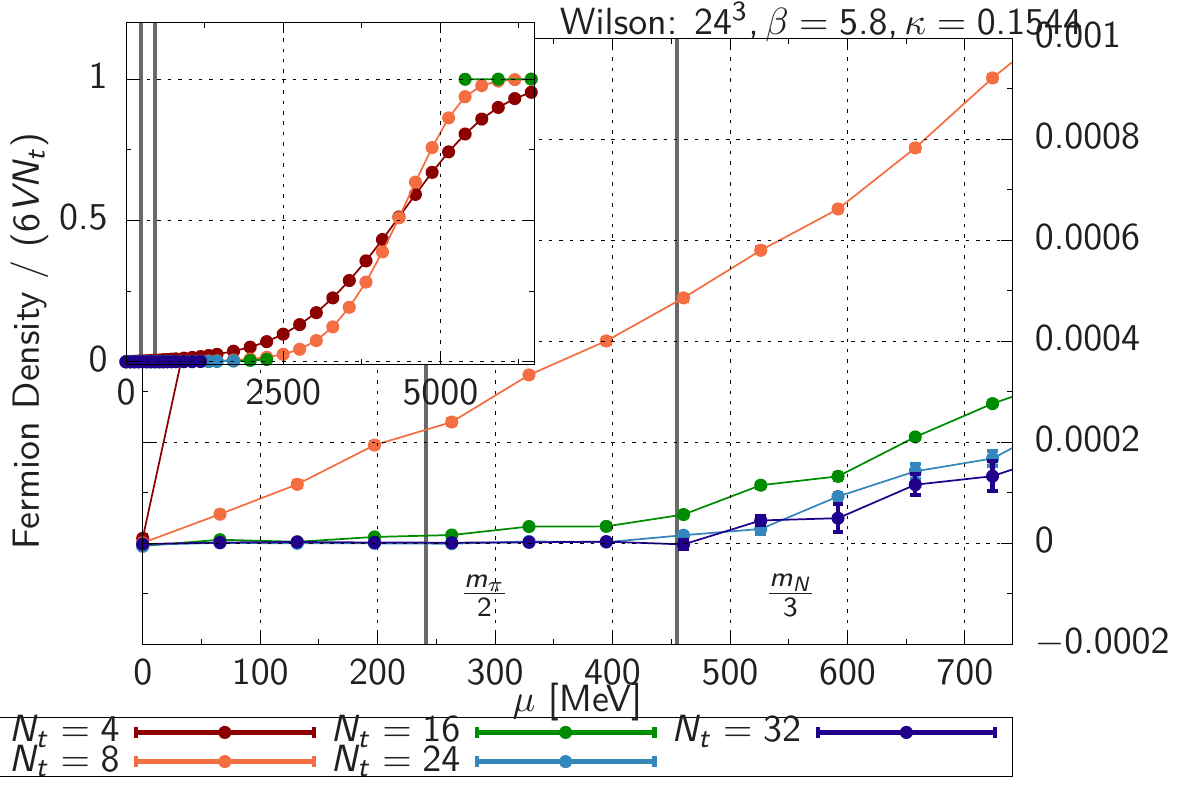}
\includegraphics[width=0.5\textwidth]{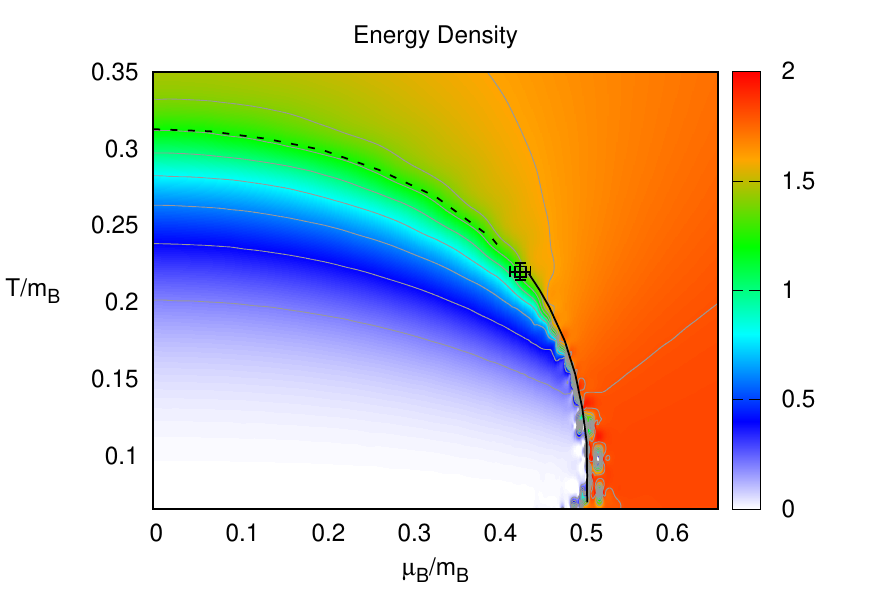}
\caption{Left: Fermion density as function of chemical potential from
  complex Langevin with $N_f=2$ Wilson fermions
  \cite{Attanasio:2021hyh}, at temperatures ranging from 100\,MeV
  ($N_t=32$) to 800\,MeV ($N_t=4$). Right: Energy density as a function of
  temperature and chemical potential in the strong coupling expansion,
  from \cite{Klegrewe:2020cnp}.  A first order nuclear gas--liquid
  transition with a critical endpoint is observed.}
\label{fig:langevin-strong}
\end{figure}

Complex Langevin methods have now reached the point where they have
been used to simulate fully dynamical QCD 
with pion masses down to $\sim500\,$MeV
\cite{Attanasio:2020spv,Ito:2020mys,Attanasio:2021hyh}.
Figure~\ref{fig:langevin-strong} shows the fermion density as a function of
the chemical potential from a simulation with $N_f=2$ Wilson fermions
with $m_\pi\approx500\,$MeV and a lattice spacing $a\approx0.06\,$fm
\cite{Attanasio:2021hyh}.  At low temperature the Silver Blaze
phenomenon is clearly seen, as the density remains zero up to
$\mu=m_N/3$ with an onset of nuclear matter beyond this point.  At
very high chemical potential we observe saturation, where all
available fermion states on the lattice are occupied.

Recent developments in strong coupling methods
\cite{Philipsen:2021vgp}, based on a character expansion of the gauge
field action, have allowed simulations to be carried out for heavy
quarks (using a hopping parameter expansion) \cite{Glesaaen:2015vtp}
as well as for light quarks (using a dual formulation)
\cite{Gagliardi:2017uag,Gagliardi:2019cpa}.  Recently, a
continuous-time formulation for staggered fermions has been developed
\cite{Klegrewe:2020cnp}.  Figure~\ref{fig:langevin-strong} shows
results using this formulation, exhibiting a first-order nuclear
gas--nuclear liquid phase transition including a critical endpoint.
The strong coupling expansion has also been applied to large-$N_c$ QCD
in the heavy dense limit \cite{Philipsen:2019qqm}, with findings
consistent with the appearance of quarkyonic matter at large $N_c$.


QCD with nonzero isospin chemical potential $\mu_I$
\cite{Brandt:2017oyy,Brandt:2019hel} and two-colour QCD (QC$_2$D)
\cite{Boz:2019enj,Astrakhantsev:2020tdl,Buividovich:2020dks,Iida:2019rah,Wilhelm:2019fvp}
do not suffer from the sign problem and are amenable to direct
simulations at large density, and may hence provide benchmarks for other
methods
\cite{Andersen:2015sma,Chao:2018czo,Adhikari:2019zaj,Khunjua:2020xws,Furusawa:2020qdz,Contant:2019lwf}.
Both feature a superfluid phase for $\mu_{I,B}>m_\pi$ characterised by
condensation of pions and scalar diquarks respectively.
The phase diagrams for both theories turn out to be remarkably
similar; in particular, the superfluid transition temperature has been
found to be almost independent of $\mu$ in both cases.  In QC$_2$D
simulations with different lattice formulations have found
consistent results for
the superfluid transition and have also confirmed the existence of a
``quarkyonic'' phase at large $\mu$, where quarks remain confined but
the bulk degrees of freedom are weakly interacting quarks.  A 
BEC r\'egime described by chiral perturbation theory has been found at
lower $\mu$ for sufficiently light quarks
\cite{Astrakhantsev:2020tdl}.  It remains an open question whether
there is a large-$\mu$ deconfinement transition at small temperature.



\begin{figure}
\centering
\includegraphics[width=0.45\textwidth]{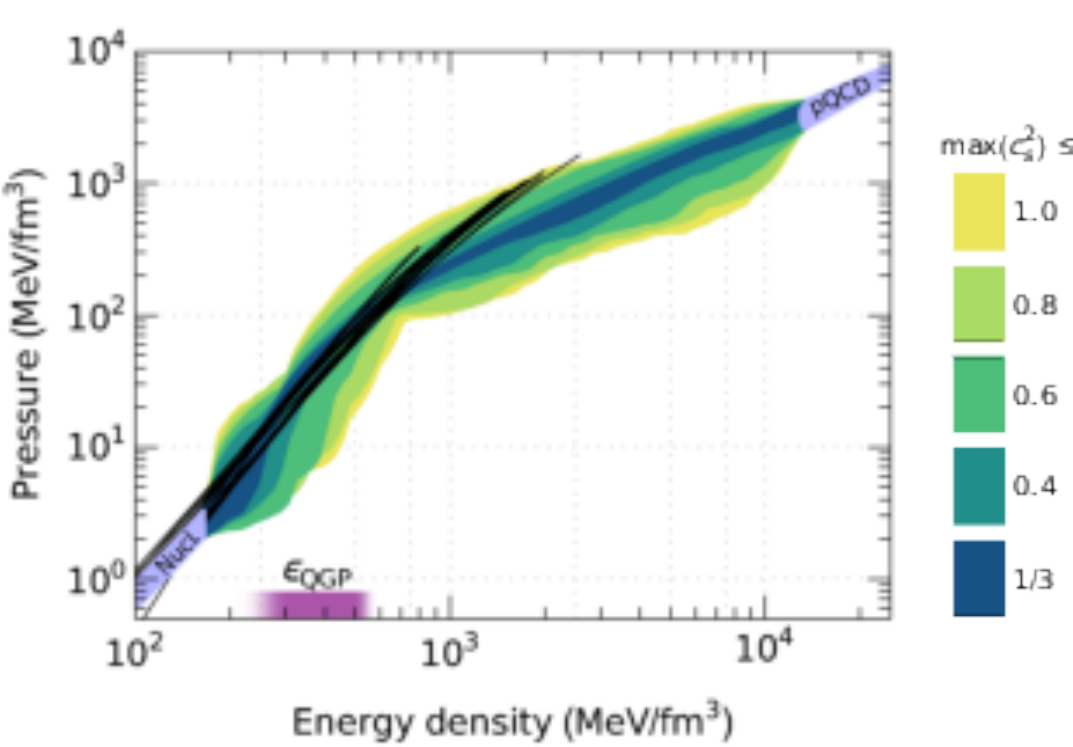}
\includegraphics[width=0.45\textwidth]{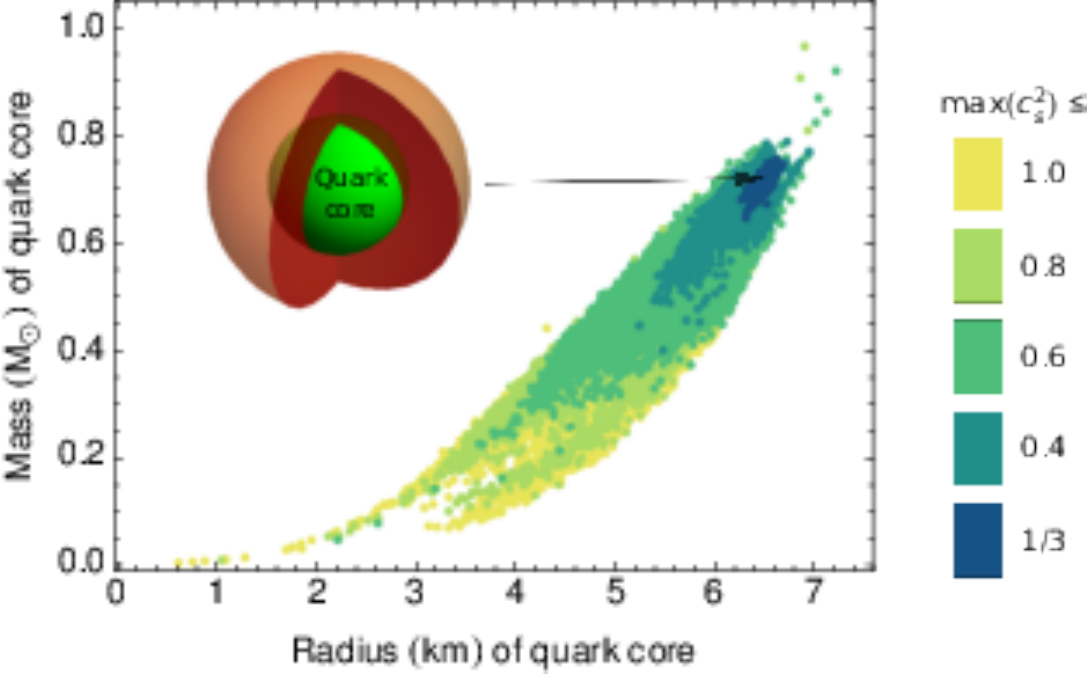}
\caption{Left: Equation of state for high-density QCD obtained by
  interpolating between a nuclear eos (bottom left hand corner) and
  perturbative QCD (top right), constrained by mass and radius data
  from compact star observations.  The kink in the eos suggests a
  transition from nuclear matter to quark matter.  Right: The size of
  the quark matter core in a $2M_\odot$ star, for the different
  interpolations shown in the left hand plot. Figures taken from
  \cite{Annala:2019puf}.}
\label{fig:star-constraints}
\end{figure}

In the absence of reliable, non-perturbative, first-principles
information on cold high-density matter, constraints can be placed by
combining information on the nuclear matter equation of state with
compact star observations.  This may be further constrained by
requiring the equation of state to match on to perturbation theory at
very high density, which has recently been computed to NNLO including
NNNLO terms \cite{Gorda:2021znl}.  This
approach has been pursued in
\cite{Annala:2019puf,Gorda:2021znl,Annala:2021gom,Greif:2020pju,Raaijmakers:2021uju}.
Specifically, the low-energy equation of state has been computed in a
chiral EFT, with the EoS at intermediate density modelled by piecewise
polytropes, $p\propto n^\gamma$.  Typically, the polytropic index
$\gamma\sim2.5$ for nuclear matter, while high-density perturbation
theory corresponds to $\gamma\in[0.5,0.8]$.  Combined with recent
observational data including from gravitational waves and the NICER
observatory \cite{Raaijmakers:2021uju} only a subset of all possible equations of
state remain.  A result of such an analysis \cite{Annala:2019puf} is shown in
figure~\ref{fig:star-constraints}, where it is
concluded that stars with a mass $2M_\odot$ will include a quark
matter core, characterised by a small polytropic index, for any EoS
that is not ruled out by observations.

\section{Hadron properties and transport}
\label{sec:spectral}

\subsection{Baryons}
\label{sec:baryons}

\begin{figure}
\centering
\includegraphics[width=0.45\textwidth]{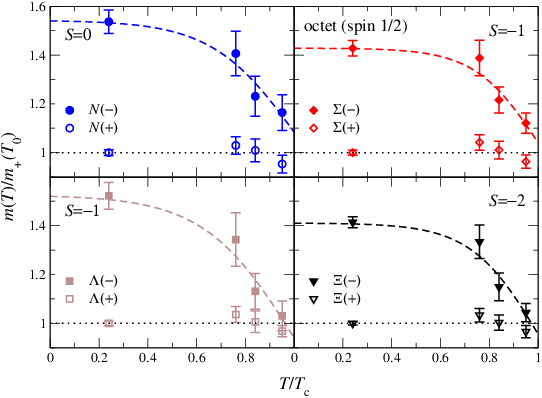}
\includegraphics[width=0.45\textwidth]{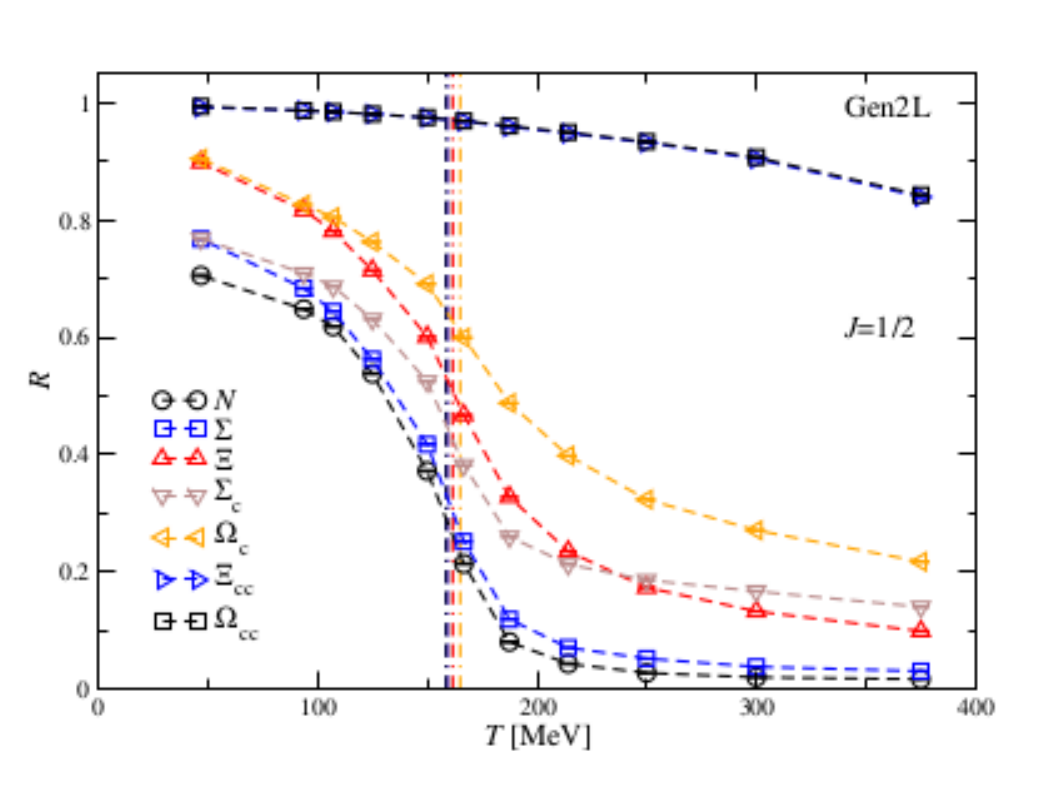}
\caption{Left: Masses of positive and negative parity octet baryons
  below the deconfinement transition, from \cite{Aarts:2018glk}.
  Right: The parity-doubling $R$ parameter \eqref{eq:R-baryon} for
  light, strange and charmed baryons.}
\label{fig:baryons}
\end{figure}

The FASTSUM collaboration has studied the properties of baryons at
high temperature
\cite{Aarts:2015mma,Aarts:2017rrl,Aarts:2018glk,Aarts:2020vyb}, and in
particular the emergent degeneracy between positive and negative
parity states.  Since positive and negative parity are
states encoded in the same correlator (as forward and backward moving
states respectively), the parameter
\begin{align}
  R &=
\frac{\sum\limits_{n=0}^{\beta/2-1}R(\tau_n)/\sigma^2(\tau_n)}
     {\sum\limits_{n=0}^{\beta/2-1}1/\sigma^2(\tau_n)}\,;&
 R(\tau) &= \frac{G(\tau)-G(\beta-\tau)}{G(\tau)+G(\beta-\tau)}
\label{eq:R-baryon}
\end{align}
may serve as a measure of parity doubling, with $R=0$ when positive
and negative parity states are completely degenerate.
Figure~\ref{fig:baryons} (left) shows the masses of
positive and negative parity octet baryons below the deconfinement
transition \cite{Aarts:2017rrl}: we see that while the masses
positive-parity baryons remain unchanged, their negative-parity
partners become lighter with increasing temperature with degeneracy
setting in close to $T_c$, with potential impact on hadron resonance
gas models \cite{Aarts:2018glk}.
The $R$ parameter \eqref{eq:R-baryon} 
is shown in figure~\ref{fig:baryons} (right).  It has an inflection
point which is roughly the same for all light, strange and even
singly-charmed baryons, and which coincides approximately with the
chiral transition temperature \cite{Aarts:2020vyb}.  Doubly-charmed
baryons, however, do not appear sensitive to the chiral transition.

\subsection{Heavy quarkonium}
\label{sec:heavyquarks}

Heavy quarks have long been a focus of attention, both theoretically
and experimentally, in the context of the quark--gluon plasma and
heavy-ion collisions, see for example the review
\cite{Rothkopf:2019ipj}.
\begin{figure}
\centering
\includegraphics[width=0.48\textwidth]{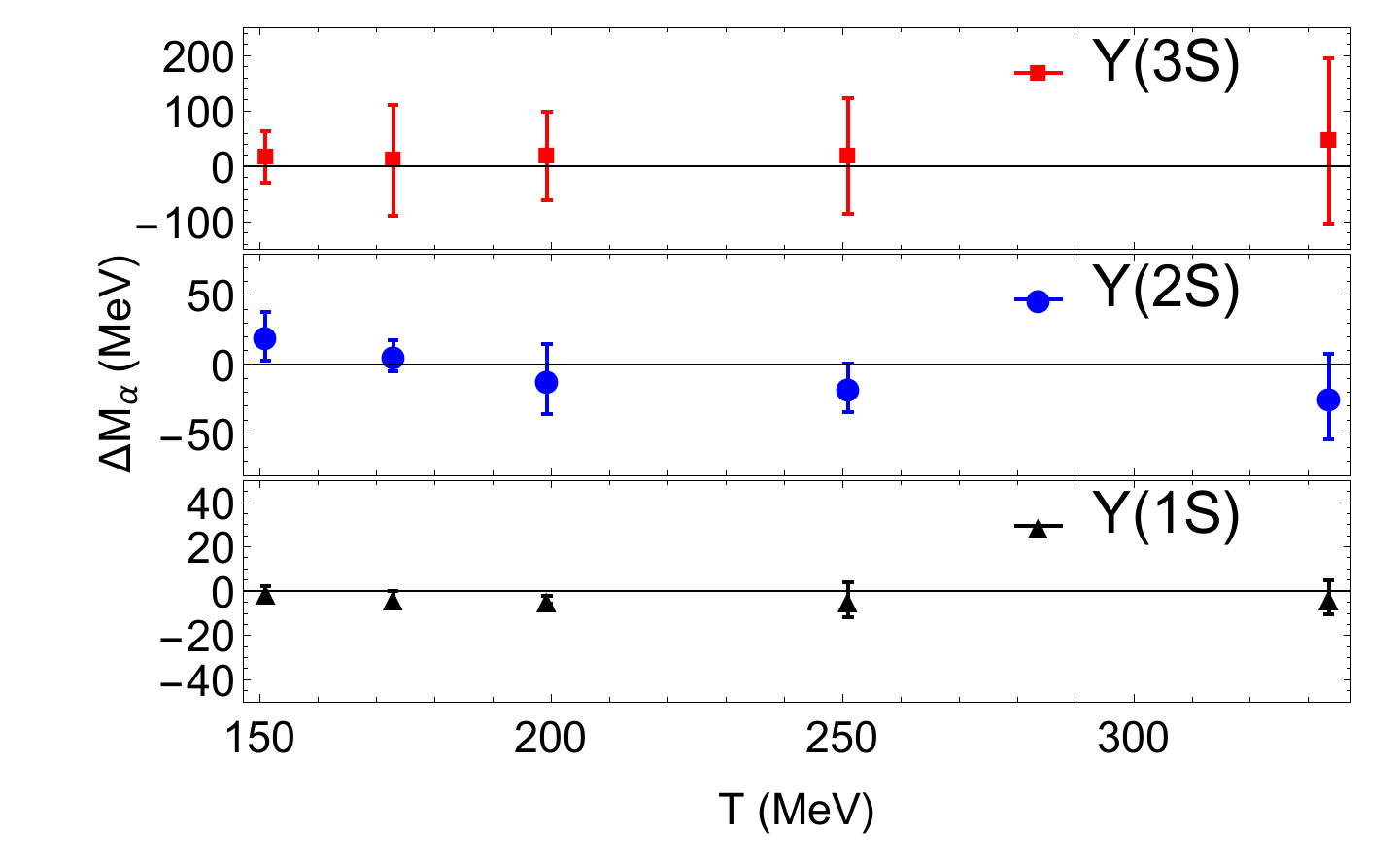}
\includegraphics[width=0.48\textwidth]{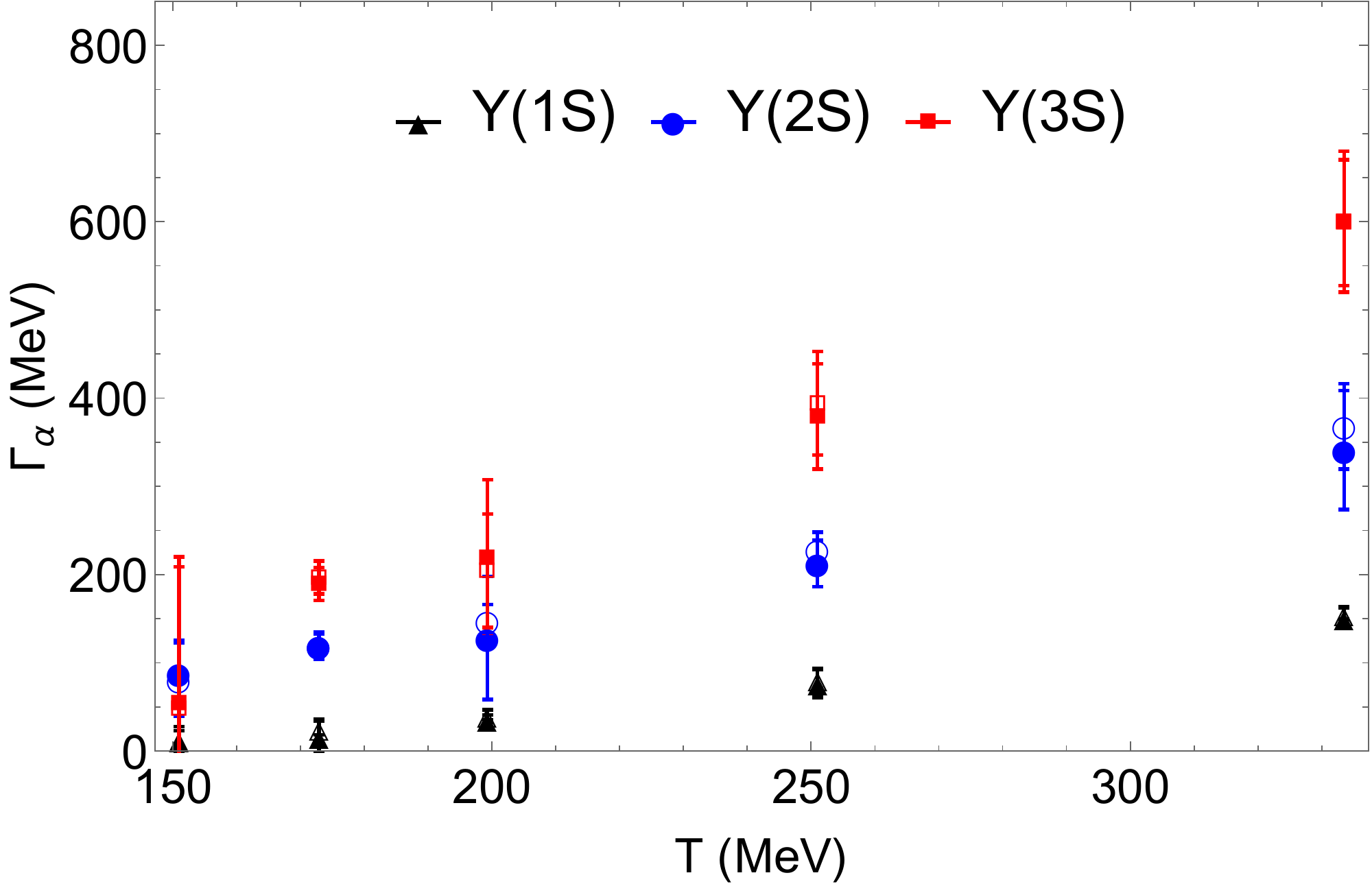}
\caption{Mass shift (left) and width (right) for the $\Upsilon$ 1S, 2S
  and 3S states, from \cite{Larsen:2019zqv}.}
\label{fig:Upsilon}
\end{figure}
In recent years there has been significant progress towards a
quantitative understanding of thermal mass shifts and widths of heavy
quarkonia, in particular in the beauty sector.  In a series of papers
\cite{Larsen:2019bwy,Larsen:2019zqv,Larsen:2020rjk} Larsen \textit{et
  al.} have employed lattice non-relativistic QCD (NRQCD) on hotQCD
ensembles with $N_f=2+1$ and a near-physical pion mass.  Using
sources optimised to reproduce the zero-temperature spectrum and
wavefunctions they have 
been able to fit the correlators to a gaussian plus low and high
energy tails.  The Bethe--Salpeter wavefunctions have also been
determined \cite{Larsen:2020rjk}.  Results for the vector ($\Upsilon$)
channel are shown in fig.~\ref{fig:Upsilon}, suggesting a very small
(negative) mass shift for the ground state and a thermal width
increasing with temperature as expected.  Similar results are found
for the P-wave $\chi_{b0}$ (scalar) channel.
It may be noted that a small negative mass shift has also been
observed by Kim \textit{et al.} \cite{Kim:2018yhk} and for charmonium
by Kelly \textit{et al.} \cite{Kelly:2018hsi}, both using the BR method
\cite{Burnier:2013nla}.

These results need to be independently confirmed using different
approaches.  The FASTSUM collaboration has employed a wide range
methods to determine the mass shift and width of the $\Upsilon$(1S)
state \cite{Spriggs:2021dsb}, including Gaussian fits, moments of
correlators, and spectral function reconstruction with linear,
Bayesian and machine learning methods.  The results are as yet
inconclusive, revealing quite different systematics for the different
methods, which need to be better understood.

The Bielefeld--CCNU collaboration has conducted a long series of
studies of quarkonium in quenched QCD using nonperturbatively improved
Wilson fermions.  Their most recent results \cite{Ding:2021ise}
suggest that there is no resonance peak above $T_c$ for $J/\psi$,
while the resonance peak for $\Upsilon$ persists only up to
$1.5T_c$.  More work is clearly needed to understand the source of
this discrepancy.


\subsection{Transport}
\label{sec:transport}

Transport properties describe the real-time, non-equilibrium
properties of the medium, and are as such not directly accessible with
the equilibrium methods of euclidean quantum field theory.  However, a
transport coefficient $\kappa$ can be related to the spectral
functions of the relevant conserved current $J_\mu^\kappa(x)$ via Kubo
relations, 
\begin{align}
\kappa &\propto \lim_{\om\to0}\frac{\rho^\kappa(\om)}{\om}\,;&\qquad
 G^\kappa_{ij}(\tau)
 &= \int
 d^3x\braket{J_i^\kappa(\tau,\vec{x})J_j^\kappa(0,\vec{0})}
  = \frac{1}{2\pi}\int\limits_0^\infty
 K(\om,\tau;T)\rho_{ij}^\kappa(\om)d\om\,,
\end{align}
where $K(\om,\tau;T)$ is a known integral kernel.  While there has
been some progress on shear and bulk viscosity \cite{Astrakhantsev:2018oue,Itou:lat21} and jet
quenching \cite{Panero:2013pla,Kumar:2020wvb}, we are still some way from obtaining
quantitatively reliable predictions for these quantities, in contrast
to the heavy quark diffusion coefficient and the electrical
conductivity, which will be reviewed below.

Two methods have been used to compute the heavy quark
diffusion coefficient from the lattice.  One method
\cite{Caron-Huot:2009ncn} uses an effective theory where the heavy
quark momentum diffusion coefficient is written in terms of a
Polyakov loop correlator with a colour-electric field insertion.  The
main benefit of this method is that the related spectral function has
a smooth behaviour as $\om\to0$, greatly reducing any uncertainties
related to determining a potentially narrow transport peak.  The main
drawback is that the correlator is inherently noisy, and noise
reduction techniques must be used to get a reasonable signal.  Recent
continuum extrapolated results from this method have been reported by
TUMQCD \cite{Brambilla:2020siz}, using a multilevel algorithm which is
applicable for pure gauge theories, and by the Bielefeld--Darmstadt
collaboration \cite{Altenkort:2020fgs,Altenkort:2021umr} using the
Wilson flow, which is also applicable in the presence of dynamical
fermions.  The results are consistent with each other and with earlier
results \cite{Banerjee:2011ra,Francis:2015daa}.

The second method uses fermionic vector correlators of charm and
beauty quarks, with the benefit that finite quark mass effects
are directly included, and the correlators are much less noisy than
the gluonic correlators in the EFT method.  The main drawback is that
it can be very difficult to reliably determine the height of a
possibly narrow transport peak.  Continuum extrapolated
results for both charm and beauty diffusion have recently been
presented by the Bielefeld--CCNU collaboration \cite{Ding:2021ise}.

\begin{figure}
\centering
\begin{minipage}[t]{0.48\textwidth}
\includegraphics[width=1.05\textwidth,clip]{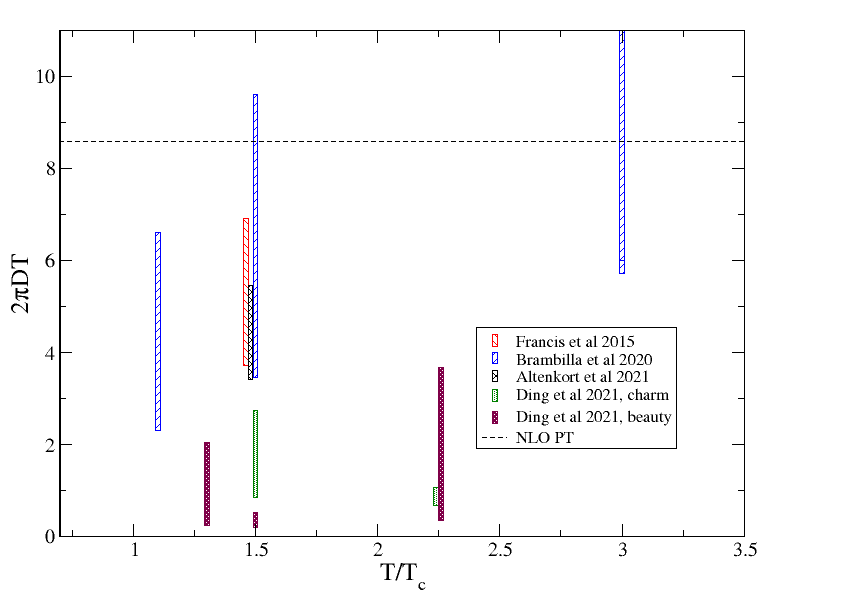}
\end{minipage}
\begin{minipage}[t]{0.48\textwidth}
\includegraphics[width=0.95\textwidth]{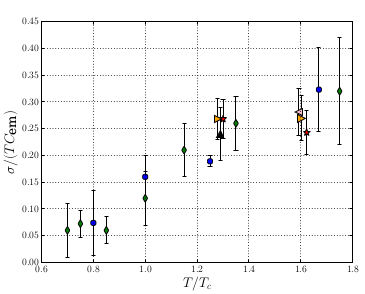}
\hspace{-0.4\textwidth}
\includegraphics[width=0.35\textwidth]{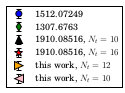}
\end{minipage}
\caption{Left: summary of continuum-extrapolated lattice results for
  the heavy quark diffusion coefficient in pure gauge theory
  \cite{Francis:2015daa,Brambilla:2020siz,Altenkort:2020fgs,Altenkort:2021umr,Ding:2021ise}.
  Hatched symbols are obtained using the EFT description; filled
  symbols are from fermionic operators.
  Right: Summary of lattice results for the electrical conductivity of
  QCD matter, from \cite{Trunin:lat21}.}
\label{fig:diff-cond}
\end{figure}

Figure~\ref{fig:diff-cond} (left) shows a comparison of lattice
results for the diffusion coefficient $D$ together with the NLO
perturbative value.  The results from each of the two methods are internally
consistent, but the results from fermionic correlators are
systematically smaller than from the effective field theory.

The electrical conductivity has been computed on the lattice by a
number of groups using various approaches (see
\cite{Aarts:2020dda} for a recent review).  In addition to earlier
results with $\order(a)$ improved Wilson fermions and heavier than
physical light quarks
\cite{Amato:2013naa,Aarts:2014nba,Brandt:2015aqk}, there are now
results using staggered fermions with physical up, down and strange
quark masses \cite{Astrakhantsev:2019zkr,Astrakhantsev:2021uii}. A
summary of these results are shown in figure~\ref{fig:diff-cond}
(right panel).  We see that there is good agreement between all the
results despite the different methods used, suggesting that we are
close to a quantitative, reliable determination of the conductivity of
the QGP and its temperature variation.  Astrakhantsev \textit{et al.}
have also computed the response of the conductivity to an external
magnetic field \cite{Astrakhantsev:2019zkr} and to a nonzero baryon
chemical potential \cite{Astrakhantsev:2021uii}.  The latter shows a
reduction in the conductivity as $\mu_B$ increases, in qualitative
agreement with a direct calculation in two-colour QCD
\cite{Buividovich:2020dks}.

\section{Summary}
\label{sec:summary}

There has been a large amount of progress on understanding the
properties of QCD at high temperature and density recently.  In
particular, we have reached the era of precision physics 
determining the chiral transition temperature $T_c$ as well as the
high-temperature equation of state, and we are within reach of
precision results for fluctuations both at zero and nonzero baryon
density.  We are also approaching a quantitative understanding of
hadron properties in the hot medium as well as some transport
properties, in particular the electrical conductivity.
At the same time, despite significant progress especially in complex
Langevin methods and constraints from compact star observations, we
are far from being able to make reliable predictions in the
low-temperature, high-density region relevant for neutron stars.  
First-principles calculations of the shear viscosity beyond theories
with gravity duals are also in their infancy.
Finally, the strong interaction has given us many surprises in the
past and continues to surprise us in the present.  It seems likely
that there will be more surprises in store also in the future.


%
\bibliography{hot,density,jis}

\begin{thebibliography}{103}

\bibitem{Fukushima:2013rx}
K.~Fukushima, C.~Sasaki, Prog. Part. Nucl. Phys. \textbf{72}, 99 (2013),
  \texttt{1301.6377}

\bibitem{Ding:2017giu}
H.T. Ding, PoS \textbf{LATTICE2016}, 022 (2017), \texttt{1702.00151}

\bibitem{Fischer:2018sdj}
C.S. Fischer, Prog. Part. Nucl. Phys. \textbf{105}, 1 (2019),
  \texttt{1810.12938}

\bibitem{Ghiglieri:2020dpq}
J.~Ghiglieri, A.~Kurkela, M.~Strickland, A.~Vuorinen, Phys. Rept. \textbf{880},
  1 (2020), \texttt{2002.10188}

\bibitem{Guenther:2020jwe}
J.N. Guenther, Eur. Phys. J. A \textbf{57}, 136 (2021), \texttt{2010.15503}

\bibitem{HotQCD:2018pds}
A.~Bazavov et~al. (HotQCD), Phys. Lett. B \textbf{795}, 15 (2019),
  \texttt{1812.08235}

\bibitem{Borsanyi:2020fev}
S.~Borsanyi, Z.~Fodor, J.N. Guenther, R.~Kara, S.D. Katz, P.~Parotto,
  A.~Pasztor, C.~Ratti, K.K. Szabo, Phys. Rev. Lett. \textbf{125}, 052001
  (2020), \texttt{2002.02821}

\bibitem{Kotov:2021rah}
A.Y. Kotov, M.P. Lombardo, A.~Trunin, Phys. Lett. B \textbf{823}, 136749
  (2021), \texttt{2105.09842}

\bibitem{Gao:2020fbl}
F.~Gao, J.M. Pawlowski, Phys. Lett. B \textbf{820}, 136584 (2021),
  \texttt{2010.13705}

\bibitem{Braun:2020ada}
J.~Braun, W.j. Fu, J.M. Pawlowski, F.~Rennecke, D.~Rosenbl\"uh, S.~Yin, Phys.
  Rev. D \textbf{102}, 056010 (2020), \texttt{2003.13112}

\bibitem{Aarts:2020vyb}
G.~Aarts et~al. (2020), \texttt{2007.04188}

\bibitem{HotQCD:2019xnw}
H.T. Ding et~al. (HotQCD), Phys. Rev. Lett. \textbf{123}, 062002 (2019),
  \texttt{1903.04801}

\bibitem{HotQCD:2012vvd}
A.~Bazavov et~al. (HotQCD), Phys. Rev. D \textbf{86}, 094503 (2012),
  \texttt{1205.3535}

\bibitem{Buchoff:2013nra}
M.I. Buchoff et~al., Phys. Rev. D \textbf{89}, 054514 (2014),
  \texttt{1309.4149}

\bibitem{Dick:2015twa}
V.~Dick, F.~Karsch, E.~Laermann, S.~Mukherjee, S.~Sharma, Phys. Rev. D
  \textbf{91}, 094504 (2015), \texttt{1502.06190}

\bibitem{Ding:2020xlj}
H.T. Ding, S.T. Li, S.~Mukherjee, A.~Tomiya, X.D. Wang, Y.~Zhang, Phys. Rev.
  Lett. \textbf{126}, 082001 (2021), \texttt{2010.14836}

\bibitem{Kaczmarek:2021ser}
O.~Kaczmarek, L.~Mazur, S.~Sharma (2021), \texttt{2102.06136}

\bibitem{Cossu:2013uua}
G.~Cossu, S.~Aoki, H.~Fukaya, S.~Hashimoto, T.~Kaneko, H.~Matsufuru, J.I.
  Noaki, Phys. Rev. D \textbf{87}, 114514 (2013), [Erratum: Phys.Rev.D 88,
  019901 (2013)], \texttt{1304.6145}

\bibitem{Brandt:2016daq}
B.B. Brandt, A.~Francis, H.B. Meyer, O.~Philipsen, D.~Robaina, H.~Wittig, JHEP
  \textbf{12}, 158 (2016), \texttt{1608.06882}

\bibitem{Tomiya:2016jwr}
A.~Tomiya, G.~Cossu, S.~Aoki, H.~Fukaya, S.~Hashimoto, T.~Kaneko, J.~Noaki,
  Phys. Rev. D \textbf{96}, 034509 (2017), [Addendum: Phys.Rev.D 96, 079902
  (2017)], \texttt{1612.01908}

\bibitem{Aoki:2020noz}
S.~Aoki, Y.~Aoki, G.~Cossu, H.~Fukaya, S.~Hashimoto, T.~Kaneko, C.~Rohrhofer,
  K.~Suzuki (JLQCD), Phys. Rev. D \textbf{103}, 074506 (2021),
  \texttt{2011.01499}

\bibitem{Braun:2020mhk}
J.~Braun, M.~Leonhardt, J.M. Pawlowski, D.~Rosenbl\"uh (QCD) (2020),
  \texttt{2012.06231}

\bibitem{Bazavov:2017xul}
A.~Bazavov, H.T. Ding, P.~Hegde, F.~Karsch, E.~Laermann, S.~Mukherjee,
  P.~Petreczky, C.~Schmidt, Phys. Rev. D \textbf{95}, 074505 (2017),
  \texttt{1701.03548}

\bibitem{Kuramashi:2020meg}
Y.~Kuramashi, Y.~Nakamura, H.~Ohno, S.~Takeda, Phys. Rev. D \textbf{101},
  054509 (2020), \texttt{2001.04398}

\bibitem{Ohno:lat21}
H.~Ohno, Y.~Kuramashi, Y.~Nakamura, S.~Takeda, \emph{{Critical endpoints in
  (2+1)- and 4-flavor QCD with Wilson-Clover fermions}} (2021), presentation at
  Lattice 2021

\bibitem{Sharma:lat21}
S.~Sharma, L.~Dini, A.~Lahiri, \emph{Chiral phase transition temperature in
  3-flavor {QCD}} (2021), presentation at Lattice 2021

\bibitem{Cuteri:2021ikv}
F.~Cuteri, O.~Philipsen, A.~Sciarra (2021), \texttt{2107.12739}

\bibitem{Rohrhofer:2017grg}
C.~Rohrhofer, Y.~Aoki, G.~Cossu, H.~Fukaya, L.Y. Glozman, S.~Hashimoto, C.B.
  Lang, S.~Prelovsek, Phys. Rev. D \textbf{96}, 094501 (2017), [Erratum:
  Phys.Rev.D 99, 039901 (2019)], \texttt{1707.01881}

\bibitem{Rohrhofer:2019qwq}
C.~Rohrhofer, Y.~Aoki, G.~Cossu, H.~Fukaya, C.~Gattringer, L.Y. Glozman,
  S.~Hashimoto, C.B. Lang, S.~Prelovsek, Phys. Rev. D \textbf{100}, 014502
  (2019), \texttt{1902.03191}

\bibitem{Rohrhofer:2019qal}
C.~Rohrhofer, Y.~Aoki, L.Y. Glozman, S.~Hashimoto, Phys. Lett. B \textbf{802},
  135245 (2020), \texttt{1909.00927}

\bibitem{Cardinali:2021mfh}
M.~Cardinali, M.~D'Elia, A.~Pasqui (2021), \texttt{2107.02745}

\bibitem{Alexandru:2019gdm}
A.~Alexandru, I.~Horv\'ath, Phys. Rev. D \textbf{100}, 094507 (2019),
  \texttt{1906.08047}

\bibitem{Alexandru:2021pap}
A.~Alexandru, I.~Horv\'ath, Phys. Rev. Lett. \textbf{127}, 052303 (2021),
  \texttt{2103.05607}

\bibitem{Bonati:2018nut}
C.~Bonati, M.~D'Elia, F.~Negro, F.~Sanfilippo, K.~Zambello, Phys. Rev. D
  \textbf{98}, 054510 (2018), \texttt{1805.02960}

\bibitem{Borsanyi:2021hbk}
S.~Borsanyi, Z.~Fodor, M.~Giordano, S.D. Katz, D.~Nogradi, A.~Pasztor, C.H.
  Wong (2021), \texttt{2108.09213}

\bibitem{Fu:2019hdw}
W.j. Fu, J.M. Pawlowski, F.~Rennecke, Phys. Rev. D \textbf{101}, 054032 (2020),
  \texttt{1909.02991}

\bibitem{Gao:2020qsj}
F.~Gao, J.M. Pawlowski, Phys. Rev. D \textbf{102}, 034027 (2020),
  \texttt{2002.07500}

\bibitem{Bonati:2015bha}
C.~Bonati, M.~D'Elia, M.~Mariti, M.~Mesiti, F.~Negro, F.~Sanfilippo, Phys. Rev.
  D \textbf{92}, 054503 (2015), \texttt{1507.03571}

\bibitem{Bellwied:2015rza}
R.~Bellwied, S.~Borsanyi, Z.~Fodor, J.~G\"unther, S.D. Katz, C.~Ratti, K.K.
  Szabo, Phys. Lett. B \textbf{751}, 559 (2015), \texttt{1507.07510}

\bibitem{Gao:2016qkh}
F.~Gao, Y.x. Liu, Phys. Rev. D \textbf{94}, 076009 (2016), \texttt{1607.01675}

\bibitem{Borsanyi:2018grb}
S.~Borsanyi, Z.~Fodor, J.N. Guenther, S.K. Katz, K.K. Szabo, A.~Pasztor,
  I.~Portillo, C.~Ratti, JHEP \textbf{10}, 205 (2018), \texttt{1805.04445}

\bibitem{Bazavov:2020bjn}
A.~Bazavov et~al., Phys. Rev. D \textbf{101}, 074502 (2020),
  \texttt{2001.08530}

\bibitem{Fu:2016tey}
W.j. Fu, J.M. Pawlowski, F.~Rennecke, B.J. Schaefer, Phys. Rev. D \textbf{94},
  116020 (2016), \texttt{1608.04302}

\bibitem{Fu:2021oaw}
W.j. Fu, X.~Luo, J.M. Pawlowski, F.~Rennecke, R.~Wen, S.~Yin (2021),
  \texttt{2101.06035}

\bibitem{Alexandru:2020wrj}
A.~Alexandru, G.~Basar, P.F. Bedaque, N.C. Warrington (2020),
  \texttt{2007.05436}

\bibitem{Langfeld:2016kty}
K.~Langfeld, PoS \textbf{LATTICE2016}, 010 (2017), \texttt{1610.09856}

\bibitem{Gattringer:2019khb}
C.~Gattringer, M.~Mandl, P.~T\"orek, Phys. Rev. D \textbf{100}, 114517 (2019),
  \texttt{1911.05320}

\bibitem{Philipsen:2021vgp}
O.~Philipsen, \emph{Strong coupling methods in {QCD} thermodynamics} (2021),
  \texttt{2104.03696}

\bibitem{Attanasio:2020spv}
F.~Attanasio, B.~J\"ager, F.P.G. Ziegler, Eur. Phys. J. A \textbf{56}, 251
  (2020), \texttt{2006.00476}

\bibitem{Ito:2020mys}
Y.~Ito, H.~Matsufuru, Y.~Namekawa, J.~Nishimura, S.~Shimasaki, A.~Tsuchiya,
  S.~Tsutsui, JHEP \textbf{10}, 144 (2020), \texttt{2007.08778}

\bibitem{Attanasio:2021hyh}
F.~Attanasio, B.~J\"ager, F.P.G. Ziegler, \emph{{With complex Langevin towards
  the QCD phase diagram}}, in \emph{{38th International Symposium on Lattice
  Field Theory}} (2021), \texttt{2111.02241}

\bibitem{Klegrewe:2020cnp}
M.~Klegrewe, W.~Unger, Phys. Rev. D \textbf{102}, 034505 (2020),
  \texttt{2005.10813}

\bibitem{Glesaaen:2015vtp}
J.~Glesaaen, M.~Neuman, O.~Philipsen, JHEP \textbf{03}, 100 (2016),
  \texttt{1512.05195}

\bibitem{Gagliardi:2017uag}
G.~Gagliardi, J.~Kim, W.~Unger, EPJ Web Conf. \textbf{175}, 07047 (2018),
  \texttt{1710.07564}

\bibitem{Gagliardi:2019cpa}
G.~Gagliardi, W.~Unger, Phys. Rev. D \textbf{101}, 034509 (2020),
  \texttt{1911.08389}

\bibitem{Philipsen:2019qqm}
O.~Philipsen, J.~Scheunert, JHEP \textbf{11}, 022 (2019), \texttt{1908.03136}

\bibitem{Brandt:2017oyy}
B.B. Brandt, G.~Endr{\H{o}}di, S.~Schmalzbauer, Phys. Rev. \textbf{D97}, 054514
  (2018), \texttt{1712.08190}

\bibitem{Brandt:2019hel}
B.B. Brandt, F.~Cuteri, G.~Endr\H{o}di, S.~Schmalzbauer, Particles \textbf{3},
  80 (2020), \texttt{1912.07451}

\bibitem{Boz:2019enj}
T.~Boz, P.~Giudice, S.~Hands, J.I. Skullerud, Phys. Rev. D \textbf{101}, 074506
  (2020), \texttt{1912.10975}

\bibitem{Astrakhantsev:2020tdl}
N.~Astrakhantsev, V.V. Braguta, E.M. Ilgenfritz, A.Y. Kotov, A.A. Nikolaev,
  Phys. Rev. D \textbf{102}, 074507 (2020), \texttt{2007.07640}

\bibitem{Buividovich:2020dks}
P.V. Buividovich, D.~Smith, L.~von Smekal, Phys. Rev. D \textbf{102}, 094510
  (2020), \texttt{2007.05639}

\bibitem{Iida:2019rah}
K.~Iida, E.~Itou, T.G. Lee (2019), \texttt{1910.07872}

\bibitem{Wilhelm:2019fvp}
J.~Wilhelm, L.~Holicki, D.~Smith, B.~Wellegehausen, L.~von Smekal (2019),
  \texttt{1910.04495}

\bibitem{Andersen:2015sma}
J.O. Andersen, T.~Brauner, W.~Naylor, Phys. Rev. \textbf{D92}, 114504 (2015),
  \texttt{1505.05925}

\bibitem{Chao:2018czo}
J.~Chao, Chin. Phys. C \textbf{44}, 034108 (2020), \texttt{1808.01928}

\bibitem{Adhikari:2019zaj}
P.~Adhikari, J.O. Andersen, Phys. Lett. B \textbf{804}, 135352 (2020),
  \texttt{1909.01131}

\bibitem{Khunjua:2020xws}
T.G. Khunjua, K.G. Klimenko, R.N. Zhokhov, JHEP \textbf{06}, 148 (2020),
  \texttt{2003.10562}

\bibitem{Furusawa:2020qdz}
T.~Furusawa, Y.~Tanizaki, E.~Itou, Phys. Rev. Res. \textbf{2}, 033253 (2020),
  \texttt{2005.13822}

\bibitem{Contant:2019lwf}
R.~Contant, M.Q. Huber (2019), \texttt{1909.12796}

\bibitem{Gorda:2021znl}
T.~Gorda, A.~Kurkela, R.~Paatelainen, S.~S\"appi, A.~Vuorinen, Phys. Rev. Lett.
  \textbf{127}, 162003 (2021), \texttt{2103.05658}

\bibitem{Annala:2019puf}
E.~Annala, T.~Gorda, A.~Kurkela, J.~N\"attil\"a, A.~Vuorinen, Nature Phys.
  \textbf{16}, 907 (2020), \texttt{1903.09121}

\bibitem{Annala:2021gom}
E.~Annala, T.~Gorda, E.~Katerini, A.~Kurkela, J.~N\"attil\"a, V.~Paschalidis,
  A.~Vuorinen (2021), \texttt{2105.05132}

\bibitem{Greif:2020pju}
S.K. Greif, K.~Hebeler, J.M. Lattimer, C.J. Pethick, A.~Schwenk, Astrophys. J.
  \textbf{901}, 155 (2020), \texttt{2005.14164}

\bibitem{Raaijmakers:2021uju}
G.~Raaijmakers, S.K. Greif, K.~Hebeler, T.~Hinderer, S.~Nissanke, A.~Schwenk,
  T.E. Riley, A.L. Watts, J.M. Lattimer, W.C.G. Ho, Astrophys. J. Lett.
  \textbf{918}, L29 (2021), \texttt{2105.06981}

\bibitem{Aarts:2018glk}
G.~Aarts, C.~Allton, D.~De~Boni, B.~J{\"a}ger, Phys. Rev. \textbf{D99}, 074503
  (2019), \texttt{1812.07393}

\bibitem{Aarts:2015mma}
G.~Aarts, C.~Allton, S.~Hands, B.~J{\"a}ger, C.~Praki, J.I. Skullerud (2015),
  \texttt{1502.03603}

\bibitem{Aarts:2017rrl}
G.~Aarts, C.~Allton, D.~De~Boni, S.~Hands, B.~J{\"a}ger, C.~Praki, J.I.
  Skullerud, JHEP \textbf{06}, 034 (2017), \texttt{1703.09246}

\bibitem{Rothkopf:2019ipj}
A.~Rothkopf, Phys. Rept. \textbf{858}, 1 (2020), \texttt{1912.02253}

\bibitem{Larsen:2019zqv}
R.~Larsen, S.~Meinel, S.~Mukherjee, P.~Petreczky, Phys. Lett. B \textbf{800},
  135119 (2020), \texttt{1910.07374}

\bibitem{Larsen:2019bwy}
R.~Larsen, S.~Meinel, S.~Mukherjee, P.~Petreczky, Phys. Rev. D \textbf{100},
  074506 (2019), \texttt{1908.08437}

\bibitem{Larsen:2020rjk}
R.~Larsen, S.~Meinel, S.~Mukherjee, P.~Petreczky, Phys. Rev. D \textbf{102},
  114508 (2020), \texttt{2008.00100}

\bibitem{Kim:2018yhk}
S.~Kim, P.~Petreczky, A.~Rothkopf, JHEP \textbf{11}, 088 (2018),
  \texttt{1808.08781}

\bibitem{Kelly:2018hsi}
A.~Kelly, A.~Rothkopf, J.I. Skullerud, Phys. Rev. \textbf{D97}, 114509 (2018),
  \texttt{1802.00667}

\bibitem{Burnier:2013nla}
Y.~Burnier, A.~Rothkopf, Phys. Rev. Lett. \textbf{111}, 182003 (2013),
  \texttt{1307.6106}

\bibitem{Spriggs:2021dsb}
T.~Spriggs et~al., \emph{A comparison of spectral reconstruction methods
  applied to non-zero temperature {NRQCD} meson correlation
  functions}, these proceedings, \texttt{2112.04201}

\bibitem{Ding:2021ise}
H.T. Ding, O.~Kaczmarek, A.L. Lorenz, H.~Ohno, H.~Sandmeyer, H.T. Shu (2021),
  \texttt{2108.13693}

\bibitem{Astrakhantsev:2018oue}
N.Y. Astrakhantsev, V.V. Braguta, A.Y. Kotov, Phys. Rev. D \textbf{98}, 054515
  (2018), \texttt{1804.02382}

\bibitem{Itou:lat21}
E.~Itou, Y.~Nagai, \emph{{QCD} viscosity by combining the gradient flow and
  sparse model-ing methods} (2021), presentation at Lattice 2021.

\bibitem{Panero:2013pla}
M.~Panero, K.~Rummukainen, A.~Sch\"afer, Phys. Rev. Lett. \textbf{112}, 162001
  (2014), \texttt{1307.5850}

\bibitem{Kumar:2020wvb}
A.~Kumar, A.~Majumder, J.H. Weber (2020), \texttt{2010.14463}

\bibitem{Caron-Huot:2009ncn}
S.~Caron-Huot, M.~Laine, G.D. Moore, JHEP \textbf{04}, 053 (2009),
  \texttt{0901.1195}

\bibitem{Brambilla:2020siz}
N.~Brambilla, V.~Leino, P.~Petreczky, A.~Vairo, Phys. Rev. D \textbf{102},
  074503 (2020), \texttt{2007.10078}

\bibitem{Altenkort:2020fgs}
L.~Altenkort, A.M. Eller, O.~Kaczmarek, L.~Mazur, G.D. Moore, H.T. Shu, Phys.
  Rev. D \textbf{103}, 014511 (2021), \texttt{2009.13553}

\bibitem{Altenkort:2021umr}
L.~Altenkort, A.M. Eller, O.~Kaczmarek, L.~Mazur, G.D. Moore, H.T. Shu,
  \emph{{Spectral reconstruction details of a gradient-flowed color-electric
  correlator}}, in \emph{{19th International Conference on Strangeness in Quark
  Matter}} (2021), \texttt{2109.11303}

\bibitem{Banerjee:2011ra}
D.~Banerjee, S.~Datta, R.~Gavai, P.~Majumdar, Phys. Rev. D \textbf{85}, 014510
  (2012), \texttt{1109.5738}

\bibitem{Francis:2015daa}
A.~Francis, O.~Kaczmarek, M.~Laine, T.~Neuhaus, H.~Ohno, Phys. Rev. D
  \textbf{92}, 116003 (2015), \texttt{1508.04543}

\bibitem{Trunin:lat21}
A.~Trunin et~al., \emph{Electromagnetic conductivity of quark-gluon plasma at
  non-zero baryon density} (2021), presentation at Lattice 2021

\bibitem{Aarts:2020dda}
G.~Aarts, A.~Nikolaev, Eur. Phys. J. A \textbf{57}, 118 (2021),
  \texttt{2008.12326}

\bibitem{Amato:2013naa}
A.~Amato, G.~Aarts, C.~Allton, P.~Giudice, S.~Hands, J.I. Skullerud,
  Phys.Rev.Lett. \textbf{111}, 172001 (2013), \texttt{1307.6763}

\bibitem{Aarts:2014nba}
G.~Aarts, C.~Allton, A.~Amato, P.~Giudice, S.~Hands, J.I. Skullerud, JHEP
  \textbf{1502}, 186 (2015), \texttt{1412.6411}

\bibitem{Brandt:2015aqk}
B.B. Brandt, A.~Francis, B.~J\"ager, H.B. Meyer, Phys. Rev. D \textbf{93},
  054510 (2016), \texttt{1512.07249}

\bibitem{Astrakhantsev:2019zkr}
N.~Astrakhantsev, V.V. Braguta, M.~D'Elia, A.Y. Kotov, A.A. Nikolaev,
  F.~Sanfilippo, Phys. Rev. D \textbf{102}, 054516 (2020), \texttt{1910.08516}

\bibitem{Astrakhantsev:2021uii}
N.~Astrakhantsev, V.V. Braguta, M.~Cardinali, M.~D'Elia, L.~Maio,
  F.~Sanfilippo, A.~Trunin, A.~Vasiliev, \emph{{Electromagnetic conductivity of
  quark-gluon plasma at non-zero baryon density}}, in \emph{{38th International
  Symposium on Lattice Field Theory}} (2021), \texttt{2110.10727}

\end{thebibliography}

\end{document}